\documentclass[12pt,preprint]{aastex}

\shorttitle{\ion{H}{1} Observations of the Local Group Dwarf WLM}
\shortauthors{Jackson et al.}

\begin{document}

\title{\ion{H}{1} Observations of the Local Group Dwarf WLM}

\author{Dale C. Jackson, Evan D. Skillman and John M. Cannon}
\affil{Astronomy Department, University of Minnesota, 116 Church St S.E., Minneapolis, MN 55455}
\email{djackson@astro.umn.edu, skillman@astro.umn.edu, cannon@astro.umn.edu}

\and

\author{St\'{e}phanie C\^{o}t\'{e}}
\affil{National Research Council of Canada, Herzberg Institute of Astrophysics, 5071 West Saanich Road, Victoria, BC V9E 2E7}
\email{stephanie.cote@nrc.gc.ca}

\begin{abstract}

We present Australia Telescope Compact Array mosaic \ion{H}{1} imaging of the Local Group dwarf irregular 
galaxy WLM. 
We find an integrated flux of 149 
Jy km s$^{-1}$ and a total \ion{H}{1} mass of (3.2$\pm$0.3)$\times10^7$ M$_\sun$. 
The major axis of the \ion{H}{1} is aligned with the stellar component of the galaxy. 
The overall \ion{H}{1} distribution is relatively smooth at our resolution and has 
a double peaked central core. One of these peaks is aligned with a region found to 
have extinction that is internal to WLM and we take this as possible evidence of a 
large molecular gas complex in the southern half of the galaxy. The other \ion{H}{1} 
peak is in close proximity to the brightest \ion{H}{2} regions. WLM's overall velocity 
field is consistent with rigid body rotation. A rotation curve is derived, and we find 
a total dynamical mass of (3.00$\pm$0.80)$\times10^8$ M$_\sun$. We also performed a wide 
field search, 38$\arcmin$ in radius, for \ion{H}{1} companions or evidence of recent 
interactions (e.g., tidal tails), and found no detections to an \ion{H}{1} mass limit 
of $M_{\rm HI}$ $>$ 8.4$\times$10$^5$ M$_\sun$. 
  
\end{abstract}



\section{Introduction}
The WLM dwarf irregular galaxy (DDO 221, UGCA 444) is a member of the Local Group. Its basic properties
are listed in Table \ref{basic}.  Since its 
discovery \citep{wol10,mel26}, there have been numerous studies of its photometric properties, stellar 
populations, star formation history, and chemical abundances \citep{van00}. Unfortunately, other than 
observations of \ion{H}{2} regions \citep{hod95}, an unsuccessful search for CO \citep{tay01}, and studies 
of its most basic infrared \citep{mel94} and radio (Huchtmeier, Seiradakis \& Materne 1981; Huchtmeier \& 
Richter 1986) properties, little has been done to study its interstellar medium.

Understanding the neutral hydrogen content in WLM is particularly important for a number of reasons. The 
first is that at a distance of 0.95 $\pm$ 0.04 Mpc \citep{dol00}, a detailed study of the stellar 
populations of WLM is within reach. This allows us to directly observe the interplay between the stars and 
the interstellar medium (ISM). From a phenomenological point of view, 
it only makes sense that we have as detailed an understanding as is possible of the most nearby galaxies to 
support assumptions about galaxies whose properties are not as easily resolvable. Also, WLM is one of only 
a handful of irregular galaxies which are close enough for detailed analyses and which are also relatively 
isolated in space. This provides us with an excellent opportunity to examine the evolution of a galaxy 
without evidence of recent interaction with its neighbors. 

One unique aspect of WLM is a discrepancy between \ion{H}{2} region and population II stellar 
abundances. A recent study by \citet{ven03} found the oxygen abundance of a WLM supergiant to be much higher 
than the nebular abundance of the sampled \ion{H}{2} regions. This is puzzling since the young stars form 
from the galaxy's ISM and thus should not have a higher metallicity than the gas. Though a larger stellar 
sample size is clearly needed to determine if the abundance offset is real, a possible explanation is that 
high metallicity stars became associated with, but did not form from, the low metallicity gas through a 
merger. If, in fact, a recent merger did occur, it may be observable through \ion{H}{1} observations.

By studying the ISM, and particularly the neutral hydrogen, of dwarf galaxies we can extract information
about recent star formation events, mergers, and the galaxy's ability to form stars in the future. 
Additionally, we can derive a rotation curve for the galaxy and determine its 
dynamical mass. As WLM is close enough to resolve into stars we can compare the derived 
mass-to-light ratio to what is expected from its stellar population.
Since \ion{H}{1} observations are the most direct method of observing the large 
scale motions of a galaxy's ISM, continued progress with the study of dark matter relies on high spatial 
and spectral resolution observations to study the kinematics of the ISM in these galaxies.

In this paper we present the highest spectral and spatial resolution study of the \ion{H}{1} in WLM
to date. Section \ref{observations} describes the ATCA mosaic observations and reduction of the data. 
In \S\ref{flux} we analyze the total \ion{H}{1} flux and compare its distribution to that of the optical. 
The \ion{H}{1} position-velocity diagrams and WLM's derived rotation curve are shown 
in \S\ref{rotation}. 
Section \ref{companions} describes our wide-field search for an \ion{H}{1} halo and for companions to WLM
and the possibility of a recent merger with a metal-poor \ion{H}{1} cloud is discussed.  
The consequences of WLM possibly being a barred galaxy are considered in \S\ref{bar} and we summarize in
\S\ref{conclusions}. 


\section{Observations and Data Reduction}\label{observations}
WLM was observed with the Australia Telescope Compact Array (ATCA) in April and May of 1996.  
The observational parameters and global \ion{H}{1} data are listed in Table \ref{observations_table}.
A mosaic of seven pointings was done producing a 75$\arcmin$ $\times$ 75$\arcmin$ 
field. The data were taken in the 375 and 750D configurations, with antenna six not in use in either, 
producing a synthesized beam size of 39$\arcsec$ $\times$ 161$\arcsec$ in RA and Dec respectively. The 
total on-source integration time was 27.25 hours. A bandwidth of 8 MHz divided into 1024 channels centered 
at $-$130 km s$^{-1}$ yielded a velocity resolution of 1.65 km s$^{-1}$.  

Calibration was done using the MIRIAD\footnote{See The MIRIAD User's Guide at 
http://www.atnf.csiro.au/computing/software/miriad/} software package. PKS 1934-638, and PKS 2354-117 were 
used as the primary and secondary flux, bandpass and amplitude/phase calibrators, respectively. Velocities
from $-$335 to $-$220 and 40 to 70 km s$^{-1}$ were chosen as representative of the continuum and UVLIN was 
used to subtract the continuum 
emission from the spectral data. Fewer channels were used at lower frequency because contamination from 
Galactic \ion{H}{1} became a problem at velocities above $-$40 km s$^{-1}$ as is shown in Figure 1.    

Each pointing was cleaned separately using CLEAN and then combined with LINMOS to create the mosaic. This 
method proved better at removing sidelobes and other artifacts than either MOSMEM or MOSSDI which treat 
the full mosaic as a single datacube. A second datacube was created by smoothing the original cube with a Gaussian 
with full width at half maximum (FWHM) of twice the beam width in RA and Dec and a Hanning function with a 
width between first nulls of four channels in velocity. Any regions not containing flux 3$\sigma$ above the 
mean of a plane containing no real emission in the smoothed cube was blanked from 
the unsmoothed cube. The cube was then visually inspected and any signal not present in three consecutive
planes was considered noise and blanked.


\section{\ion{H}{1} Flux and Distribution of WLM}\label{flux}

\subsection{The Global \ion{H}{1} Profile}

The global \ion{H}{1} emission profile and channel maps are shown in Figures \ref{global} and \ref{channel}
respectively. In particular, note the asymmetric profile in Figure \ref{global}. Often such asymmetries are 
associated with interactions. We find a total integrated \ion{H}{1} flux of 149 Jy km s$^{-1}$. Adopting a 
distance to WLM of 0.95 Mpc \citep{dol00} and using Equation 1 of \citet{rob62} (see also the discussion by 
\citet{gio88}), the total \ion{H}{1} mass is given by:  
\begin{equation}
M_{\mbox{\ion{H}{1}}} = 2.36 \times 10^5 (\frac{S_{\mbox{\ion{H}{1}}}}{\mbox{Jy km s$^{-1}$}})
(\frac{D}{Mpc})^2 M_\sun
\end{equation}
where S$_{\rm HI}$ is the source flux in units of Jy km s$^{-1}$ and D is the distance in Mpc. We find a 
total \ion{H}{1} mass of (3.2$\pm$0.3)$\times10^7$ M$_\sun$ which is smaller than the single-dish 
measurements of \citet{huc81} (5.3$\times10^7$ M$_\sun$) and the recent \citet{kor04} value 
(4.7$\times10^7$ M$_\sun$) from the HIPASS, when corrected to the distance value adopted here. It is not 
surprising that we find a smaller value since missing short spacing information leaves us insensitive to 
large scale emission, which is obvious since \citet{huc81} report an \ion{H}{1} extent of 45$\arcmin$, 
which is three times the value we find. Note the asymmetric profile in Figure \ref{global}. Often such 
asymmetries are associated with interactions, and this presented additional motivation for a wide-field 
search for \ion{H}{1}.

\subsection{The \ion{H}{1} Distribution}

Figure \ref{fire_maps} shows the total \ion{H}{1} intensity map superposed on the stacked U image from the 
Local Group Survey 
of \citet{mas02}. The distribution is relatively smooth at our resolution and we are not
resolving the structures left by supernovae and stellar winds that we would expect from a galaxy with a 
high current star formation rate \citep{dol00}. \ion{H}{1} emission can be seen to extend much past the 
optical body which is typical of irregular galaxies; 
the \ion{H}{1} exponential scale length of the major axis is 9$\arcmin$, compared to 3.3$\arcmin$ in the 
optical \citep{mat98}. Nearly elliptical isointensity contours are observed from the outermost regions 
of the galaxy to within approximately 2$\arcmin$ in RA and 5$\arcmin$ in Dec of the center and the 
position angle of the major axis matches fairly well with that of the optical. In the northern half 
of the galaxy the \ion{H}{1} emission follows the midplane of the galaxy with a slight concave warp 
toward the west. The stellar distribution in the north has a similar shape but with a much more pronounced
warp, thus the bulk of the \ion{H}{1} emission in the north does not closely follow the stellar 
light distribution. In the 
southern half of the galaxy the \ion{H}{1} flux is very well aligned with the stars out to the edge of the 
optical emission and then shifts slightly eastward. 

Within 2$\arcmin$ in RA and 5$\arcmin$ in Dec of the galaxy center, an offset double peak emerges. 
The origin of this double peak structure is not obvious. It appears as though there are two separate 
systems, with peaks that are offset by approximately 1$\arcmin$ in RA and 3$\arcmin$ in 
Dec and with the western-most of the two peaks at a radial velocity 20 km s$^{-1}$ higher (more negative) 
than the eastern peak. This is reminiscent of the Sculptor Magellanic dwarf ESO 245-G005 (A143; 
C\^{o}t\'{e}, Carignan \& Freeman 2000) which also has a multi-peaked flux distribution with one of its 
peaks at an offset radial velocity from the rest of the galaxy. The implications of the double-peaked 
feature are discussed in \S\ref{companions}.

\subsection{Comparison With Optical Data}\label{optical}
Further inspection of the double peak region shows additional new findings. Analysis of isophotes in B 
and V from \citet{abl77} shows what could be patchy regions of extinction that subtend much of the southern 
part of WLM. An analysis of U, B, and V images from the Local Group Survey \citep{mas02} identifies
regions of higher values of both ($\ub$) and ($\bv$) in the southern part of the galaxy 
(relative to regions in the north; see Figure \ref{UB}). 
These regions are found in fingers that run between \ion{H}{2} regions, 
from the outer to the inner galaxy. 
Of the two peaks in emission we observe, the eastern 
one is roughly aligned with one of these extinction regions, while the western peak lies just north of a 
large complex of \ion{H}{2} regions. 
The all-sky Galactic extinction map produced by Schlegel, Finkbeiner, \& Davis (1998) shows E(\bv) at levels of only 0.02 to 0.03 
magnitudes; thus we assume this extinction to be internal to WLM (although the Schlegel et al.\  
observations do not resolve the galaxy so we can't be sure from their maps that there is no 
small scale Galactic extinction). 
Additionally, the \ion{H}{2} regions mapped by \citet{hod95} all lie either adjacent to or along the line 
of sight to these extinction regions. A possible explanation of this extinction is the 
presence of a large molecular gas complex in the southern half of WLM. The only search for any type of 
molecular gas was that of \citet{tay01}. Their CO observations resulted in a non-detection, however they 
point out that in low metallicity systems like WLM there is evidence for a high CO to H$_2$ conversion 
factor (Taylor, Kobulnicky, \& Skillman 1998). Also, their beam size was small enough that they were only able to observe a relatively small 
fraction of the galaxy, with those observations focusing on parts of the galaxy near the \ion{H}{2} regions
observed by \citet{hod95}. Since CO had been found near \ion{H}{2} regions in other Local Group dwarf 
irregulars, this was an excellent place to begin their search; however these regions, in large 
part, do not coincide with the regions we find to have the highest extinction. Thus, it is still possible to 
have a substantial amount of molecular hydrogen in the galaxy without any CO detected. 
We also inspected radio continuum maps to search for emission from embedded super-star clusters, which we 
might expect to observe since WLM is a late type star-forming galaxy, but none were detected.


\section{\ion{H}{1} Position-Velocity Diagrams and the Rotation Curve}\label{rotation}
Except in the far northern and southern parts of the galaxy, the isovelocity contours of WLM (shown overlaid
on the total \ion{H}{1} intensity map in Figure \ref{velocity}) are relatively parallel, which is consistent 
with regular solid body rotation. In the far north and south the contours take on the characteristic shape 
of differential rotation showing where the rotation curve begins to flatten. Overall, the velocity field is 
fairly symmetric, with some deviations we are unable to resolve along the south-eastern edge of the galaxy. This 
largely symmetric velocity distribution with slight departures is typical of dwarf irregular galaxies 
\citep{ski96}.  One noticeable asymmetry is the offset in velocity between the western and eastern sides
of the galaxy.  Although the major axis of the galaxy has a nearly N-S orientation, the isovelocity 
contours do not align with constant declination and show a difference of roughly 10 km s$^{-1}$ from one
side of the galaxy to the other.

In Figure \ref{pv} we show position-velocity diagrams for slices running 
from north to south through the east emission peak, the nominal galaxy center, and the west emission peak, 
respectively. Like the \ion{H}{1} flux distribution, the position-velocity diagrams describe a mostly 
uniform rotating disk galaxy, with some subtleties.  
First, the \ion{H}{1} emission from the eastern slice is offset from that of the western slice by 
approximately 20 km s$^{-1}$. 
Second, the eastern slice shows a bifurcation in \ion{H}{1} velocity near the center of the galaxy.
This can also be seen in the central and western slices, but is most prominent in the east.
Because of our relatively poor resolution in declination, it is not clear
what role beam smearing is playing in our perception of this feature - it could be due to a single
\ion{H}{1} cloud at a discrepant velocity or it could really be an extended feature as it appears in
Figure \ref{pv}.
This structure may be indicative of a bar in WLM and is discussed further in \S\ref{bar}. 

It is also possible that a gaseous blowout created by a stellar association is responsible for 
displacing gas north-east of the galaxy center and creating the multi-component position-velocity 
diagrams.  
Holes are relatively common in the disks of dwarf irregular galaxies \citep{ski96} and sometimes they
are directly associated with \ion{H}{1} at discrepant velocities indicative of blow-out (e.g., 
IC 10 - \citet{ss89}, \citet{wm98}; NGC 1569 - \citet{id90}, \citet{si02}; Ho II - \citet{pwbr92}).  
For a disk seen edge-on, it may not be possible to see the holes, and the only evidence of blowout would 
be gas at discrepant velocities (e.g., NGC 625 - \citet{cmsc04}).  
However, no H$\alpha$ is observed in this region and there is no evidence of a young
stellar association or cluster. Thus, if a blowout did occur, it would have to be a relatively old feature
(i.e., $>$ 10-20 Myr).
Additionally, the offset between the velocities seen in the western and eastern parts of the
galaxy are continuous over the most of the disk, so a localized blowout event seems 
an unlikely
explanation for the offset.  Ultimately, higher resolution and sensitivity \ion{H}{1} observations and
deeper optical imaging should be able to determine the nature of the discrepant velocity \ion{H}{1}.  

Rotation curves were created by fitting a tilted-ring model \citep{beg87} to the velocity field shown 
in Figure \ref{velocity}. The velocity field was divided into concentric rings, each 162$\arcsec$ in 
width, and six parameters (systemic velocity, rotational velocity, position angle, inclination, and x and 
y kinematical center) were iteratively fitted to best model the observed radial velocity. Rotation curves
were created using only the approaching half of the galaxy, only the receding half of the galaxy, and 
the entire galaxy. The systemic velocity and galaxy center were determined first, beginning with values 
from the global profile and optical data, respectively, then fixed for all rings. The inclination was fit 
to the optical value of 69$\degr$ because it was not well-constrained by the \ion{H}{1} data. The position 
angle was allowed to vary, yet on both sides of the galaxy it only changed by approximately 4\degr. 

For the final rotation curve the position angle was fixed at 181\degr, which was the average value when 
allowed to vary, leaving the rotational velocity the only parameter free to vary. Errors 
were determined by adopting the larger of either the formal errors from the fits or the difference
between the velocity with the position angle fixed and the velocity with the position angle free.
A significant limitation to our rotation curve analysis was that our beam was elongated in the same 
direction as the galaxy, leaving us few data points for the final rotation curves. Because of this, we 
will not attempt a detailed mass-modeling of the galaxy, instead leaving that task for higher-resolution
studies.

The morphology of the rotation curves are typical of a disk system, showing solid-body rotation in the inner
galaxy and differential rotation in the outer.  
For the total dynamical mass determination we will adopt the last measured data point (38 $\pm$ 5.0 km 
s$^{-1}$ at 0.89 kpc) from the rotation curve using the entire galaxy. This maximum velocity is typical 
of galaxies with similar absolute B magnitudes \citep{cot00}.  Assuming a spherically symmetric 
dark matter distribution, this yields a total dynamical mass of (3.00$\pm$0.80)$\times$10$^8$ M$_\sun$.
The (M/L$_B$)$_{dyn}$ value at the last measured point is 5.9, which is also typical of dwarf irregular galaxies 
at this radius. It should again be noted however, that due to our lack of sensitivity to large-scale 
structures, we are missing a significant amount of \ion{H}{1} at large radii and thus are certainly not 
measuring the entire dynamical mass of the galaxy. 


\section{Wide-Field Search for \ion{H}{1}}\label{companions}
After \citet{min96} suggested the existence of an extended stellar halo in WLM, mosaic imaging was proposed 
to search for an associated \ion{H}{1} halo. This search took on greater importance after a recent study by 
\citet{ven03} found a supergiant with an oxygen abundance 0.68 dex ($> 3\sigma$) higher
than the nebular abundance from sampled \ion{H}{2} regions \citep{hod95}. This is a serious problem since
stars form out of, and thus should not have higher abundances than the nearby ISM, so the possibility 
existed that
this could be the signature of a recent merger with a low-metallicity \ion{H}{1} cloud. It is interesting to
note that in ESO 245-G005, which has a multi-component flux distribution and velocity field, like WLM, was
found by \citet{mil96} to have a strong abundance gradient along its bar which is not typical of this type
of galaxy. \citet{cot00} postulate that this could be due to the accretion of an \ion{H}{1} cloud of
different metallicity or that the entire galaxy is the result of a recent merger that has not yet been
able to completely mix its metals. Their argument is strengthened by the observations of \citet{mil96},
\citet{tay95}, and others who have found that a large fraction of irregular and \ion{H}{2} galaxies have
\ion{H}{1} companions. Additionally, WLM's \ion{H}{1} global profile shows strong asymmetry, which is often 
associated with interactions in spiral galaxies.

\citet{ven03} were skeptical that the accretion of an \ion{H}{1} cloud could be responsible for the
abundance anomaly because within the last 10 Myr (the lifetime of a supergiant) a $>10^6$ M$_\sun$ cloud
would had to have been accreted to dilute the ISM sufficiently to produce the observed abundances. Because
the overall \ion{H}{1} distribution is smooth and the rotation curves for the approaching and receding
portions of the galaxy are nearly identical, with no tidal structures observed, we see no compelling
kinematic evidence of a recent interaction or merger. Additionally, we created  
a cube with three times lower resolution than the full resolution cube to search for extended low surface
brightness features like we might expect from a low-mass companion. A plane by plane search revealed no
such companions to a three sigma mass limit of 8.4$\times10^5$ M$_\sun$ (assuming a velocity width of 15 km 
s$^{-1}$), 
although because WLM is a bright source resolved by our beam we were unable to completely remove the first 
sidelobe, who's peak lies approximately 7.5$\arcmin$ east and west of the galaxy center. This leaves us 
less sensitive to any companions which may exist in this region. Finally, \citet{min96} postulated the 
existence of a stellar halo from deep $\it$V and $\it$I photometry. Their data showed the red stars to be 
less 
centrally distributed than the blue stars and attributed the difference in stellar population between the 
inner and outer part to a flattened population II halo extending to 2 kpc from the galaxy center. 
More recently however, \citet{bat04} performed a survey of C stars in WLM and found no evidence to support
the \citet{min96} conclusion. In this investigation we see no evidence of any extended \ion{H}{1} emission
which would be consistent with a gaseous halo.


\section{Is WLM a Barred Galaxy?}\label{bar}
Perhaps a better explanation for the somewhat peculiar flux distribution and irregular structure in the
position-velocity diagrams is the presence 
of a bar. Simulations of gas dynamics in barred spiral galaxies \citep{rob79} have shown that a bar and its 
associated shocks are capable of depleting gas from the inner parts of the galaxy and piling it up near 
where the end of the bar meets the spiral arms. This is born out in observations of many barred galaxies 
with giant \ion{H}{2} regions found at the ends of their bars. This is particularly true of Magellanic 
irregulars. \citet{elm80} found that 30-50\% of barred irregular galaxies in their sample had large 
peripheral \ion{H}{2} regions and that these regions are usually near one end of the bar. Interestingly, 
WLM's largest \ion{H}{2} region appears near the southwest edge of the stellar distribution. However, 
whether this region is actually at the periphery of the galaxy and furthermore whether or not WLM is a 
barred system may be impossible to determine because of its high inclination angle (69\degr). However, 
if a bar 
does exist, it might be capable of displacing enough gas to produce a multi-peaked flux distribution as 
well as inducing non-circular motions in the gas to create the observed peculiarities in the 
position-velocity diagrams. This hypothesis does not explain the abundance anomaly, but again, more 
spectroscopic observations are needed to resolve this question. 


\section{Results and Conclusions}\label{conclusions}
We have presented ATCA neutral hydrogen mosaic imaging of the dwarf irregular galaxy WLM. We find a total 
\ion{H}{1} flux of 149 Jy km s$^{-1}$ and, adopting a distance of 0.95 $\pm$ 0.04 Mpc, a total \ion{H}{1} 
mass of (3.2 $\pm$ 0.3)$\times$10$^7$ M$_\sun$. This mass is significantly lower than the single dish 
measurement of \citet{huc81} which is expected since we are insensitive to large scale emission due to short 
spacing information.

At our resolution the \ion{H}{1} distribution is smooth and extends past the optical distribution by a 
factor of approximately three which is typical of this type of galaxy. In the northern half
of the galaxy the \ion{H}{1} exhibits a slight warp toward the west though it is not as pronounced 
as that of the stars, so the bulk of the \ion{H}{1} does not closely trace the stellar light distribution. 
In the southern
half of the galaxy the \ion{H}{1} is aligned with the stars out to the edge of the optical emission
then shifts somewhat eastward. In the center of WLM we detect a double peaked core not 
previously resolved. The peaks are separated by approximately 1$\arcmin$ in RA and 3$\arcmin$ in Dec with 
the northern peak at a radial velocity 20 km s$^{-1}$ higher than the southern and while the 
northern peak is smooth in position-velocity space the southern peak is confined to between
$-$110 and $-$130 km s$^{-1}$. The stellar halo proposed by \citet{min96} led us to 
conduct a wide field search, 38$\arcmin$ in radius, for an associated gaseous halo, but we observe 
the \ion{H}{1} to fall smoothly into the background and thus see no evidence for such a halo in \ion{H}{1}. 

Inspection of the central core of WLM from a previous optical study \citep{abl77} as well as our own 
analysis of recently released data \citep{mas02} show what appear to be regions of significant extinction 
in the southern half of the galaxy. Because the Galactic extinction maps of \citet{sch98} show very little 
extinction along this line of sight and because we find variable extinction values within the galaxy, we 
conclude that the extinction is internal to WLM and produced by a complex of molecular clouds which are 
outlined by previously mapped \ion{H}{2} regions \citep{hod95}.  

The isovelocity contours of WLM are nearly parallel over most of the galaxy which is 
consistent with solid body rotation. In the far north and south the contours take on the 
characteristic shape of differential rotation showing where the rotation curve flattens. 
The velocity field is mostly symmetric, with some deviation along the south-eastern edge of the galaxy, 
which we are unable to resolve.

We do know that WLM has a double peaked flux distribution. We also see some irregularity in
the position-velocity diagrams which, when combined with the abundance anomaly of \citet{ven03} present 
an interesting puzzle since we see no kinematic evidence of a recent merger. One possible explanation for 
these observations is that WLM is a barred galaxy. Bars in dwarf irregulars have been observed to deplete 
gas from the inner parts of the galaxy and deposit it near the end of the bar. Typically giant \ion{H}{2} 
regions are found near where the bar meets the spiral arm. This would explain both the unusual flux 
distribution and velocity structures, however WLM's high inclination may make this determination 
impossible.  Only higher resolution studies of WLM will be able to shed light on what the 
true cause of the multi-peaked core is, and if WLM does have a multi-component velocity 
field.


\acknowledgments
We wish to acknowledge conversations with Eric Wilcots, and, in particular, his encouragement to 
investigate the possibility of blowout to explain the discrepant velocity gas.  
D.C.J.\ wishes to thank B.J.J., N.L.J., W.L.J., E.W.J.\ and the 309ers for their support.
E.D.S.\ is grateful for partial support from NASA LTSARP grant NAG5-9221 and the University of 
Minnesota. 
J.M.C.\ is supported by NASA Graduate Student Researchers Program (GSRP) Fellowship NGT 5-50346.
The ATNF is funded by the Commonwealth of Australia for operation as a National Facility managed by CSIRO. 
This research has made use of NASA's Astrophysics Data System Bibliographic Services and the NASA/IPAC 
Extragalactic Database (NED) which is operated by the Jet Propulsion Laboratory, California Institute of 
Technology, under contract with the National Aeronautics and Space Administration.


\clearpage
\begin{deluxetable}{lcc}
\tablecaption{Basic Properties of WLM}
\tablewidth{0pt}
\tablehead{
\colhead{Quantity} & \colhead{Value} & \colhead{Reference}}
\startdata
Right ascension, $\alpha$(2000) & 00 01 57.8 & 1\\
Declination, $\delta$(2000) & $-$15 27 51 & 1\\
Heliocentric velocity, V$_\sun$ (km s$^{-1}$) & $-$120$\pm4$ km s$^{-1}$ & 5\\
Heliocentric velocity, V$_\sun$ (km s$^{-1}$) & $-$130 km s$^{-1}$ & 8\\
Distance, D (Mpc) & 0.95 $\pm$ 0.04 & 7\\
Morphological Type & Ir IV-V & 6\\
Total \ion{H}{1} mass (M$_\sun$) & 5.3$\times10^7$ & 3\\
Full line width at half power (km s$^{-1}$) & 20 & 4\\
Inclination angle (degrees) & 69 & 2\\
Position angle (degrees) & 181 & 8\\
Rotational velocity (km s$^{-1}$) & 38 & 8\\
Conversion factor (pc/arcmin) & 276 & 8\\
\enddata
\label{basic}
\tablerefs{(1) \citet{gal75}. (2) \citet{abl77}.  (3) \citet{huc81}. (4) \citet{huc86}. (5) \citet{dev91}. 
(6) \citet{van94}. (7) \citet{dol00}. (8) This work.}
\end{deluxetable}   
\clearpage

\clearpage 
\begin{deluxetable}{lc}
\tablecaption{ATCA \ion{H}{1} Observations of WLM}
\tablewidth{0pt}
\tablehead{
\colhead{Quantity} & \colhead{Value}}
\startdata
Dates of observations & 1996 April 12,13 May 21,22,23\\
Total bandwidth (MHz) & 8\\
Channel width & 1.65 km s$^{-1}$\\
Antenna baselines 375 Configuration (meters) & 31, 61, 92, 122, 184, 214, 245, 276, 337, 459\\
Antenna baselines 750D Configuration (meters) & 31, 107, 184, 291, 398, 429, 582, 612, 689, 719\\   
Synthesized beam FWHM (arcsec) & 39 $\times$ 161\\
Conversion factor (1 Jy beam$^{-1}$)(K) & 95.59\\
Total \ion{H}{1} flux (Jy km s$^{-1}$) & 149\\
Total \ion{H}{1} mass (M$_\sun$) & (3.2$\pm$0.3)$\times10^7$\\ 
Total dynamical mass (M$_\sun$) & (3.0$\pm$0.8)$\times10^8$\
\enddata
\label{observations_table}
\end{deluxetable}
\clearpage
   

\begin{figure}
\begin{center}
\caption{Spectrum of WLM from data taken on April 12, 1996. Velocities from $-$335 to $-$220 and 40 to 70 km 
s$^{-1}$ were chosen to perform continuum subtraction, since they avoid contamination by Galactic 
\ion{H}{1}. The upper and lower curves are for the yy and xx polarizations, respectively. See 
\S\ref{observations}}.
\end{center}
\end{figure}

\begin{figure}
\begin{center}
\caption{\label{global} The high-velocity resolution global \ion{H}{1} emission profile of WLM. Note the 
strong asymmetry. This is discussed in more detail in \S\ref{rotation}.}
\end{center}
\end{figure}

\begin{figure}
\begin{center}
\caption{\label{channel} Channel maps for WLM. Contours are 2.5, 5, 10, and 20 $\sigma$ above the mean of 
the background ($\sigma$ = 18 mJy beam$^{-1}$).The synthesized beam size is shown (filled) in the lower 
left corner.}
\end{center}
\end{figure}

\begin{figure}
\begin{center}
\caption{\label{fire_maps} (a): The total \ion{H}{1} intensity map of WLM superposed on the Local Group 
Survey U image \citep{mas02}. Contours are from 25 to 95 percent of the peak column density 
(2.16$\times10^{21}$ cm$^{-2}$) in increments of 10 percent. 
(b): The same \ion{H}{1} contours superposed on the ($\ub$) image. The image has been blanked 
against the first moment map to remove low signal to noise data. The synthesized beam size for the 
\ion{H}{1} data is shown in the lower left corner.
(c): The continuum subtracted H$\alpha$ image from the Local Group Survey, again with our 
\ion{H}{1} contours overlaid. Note the two \ion{H}{1} peaks and their relationship to the bright stellar
clusters and \ion{H}{2} regions.}
\end{center}
\end{figure}

\begin{figure}
\begin{center}
\caption{\label{UB} ($\ub$) image using data from \citet{mas02}. A contour is added to guide the reader's 
eye to redder (darker) regions in the central part of the galaxy. These regions are likely due to extinction 
internal to WLM associated with molecular gas. See \S\ref{optical}. }
\end{center}
\end{figure}

\begin{figure}
\begin{center}
\caption{\label{velocity} Total intensity map for WLM with the velocity field overlaid. Contours range from 
$-$155 km s$^{-1}$ (at top) to $-$100 km s$^{-1}$ (at bottom) in increments of 5 km s$^{-1}$. The three 
lines at top and bottom show which cuts were used to create the position-velocity diagrams in Figure 
\ref{pv}. Note the systematic shift in velocity from the east the west.}
\end{center}
\end{figure}

\begin{figure}
\begin{center}
\caption{\label{pv} Position-velocity diagrams for cuts running north to south through the easternmost of 
the two emission peaks (a), the nominal galaxy center (b), and the western-most emission peak(c). There 
appears to be a two component nature which we discuss in \S\ref{rotation}.} 
\end{center}
\end{figure}

\begin{figure}
\begin{center}
\caption{\label{all_rotation} Rotation curves for WLM. Triangles are the rotation curve created using only 
the receding (south) side of the galaxy, upside-down triangles are for only the approaching (north) side 
of the galaxy, and filled circles are for the full galaxy. All rotation curves were done with fixed 
systemic velocity of $-$127 km s$^{-1}$, position angle of 181$\degr$, and inclination angle of 69$\degr$.
Note the good agreement between the independent measurements from the two sides of the galaxy and the 
transition from solid-body to differential rotation at approximately 2 kpc.}
\end{center}
\end{figure}


\begin{thebibliography}{}
\bibitem[Ables \& Ables(1977)]{abl77} Ables, H.~D.~\& Ables, P.~G. 1977, \apj, 34, 245
\bibitem[Barnes et al.(2001)]{bar01} Barnes, D.~G. et al. 2001, \mnras, 322, 486
\bibitem[Battinelli \& Demers(2004)]{bat04} Battinelli, P.~\& Demers, S. 2004, \aap, 416, 111
\bibitem[Begeman(1987)]{beg87} Begeman, K. 1987, Ph.D. thesis, Univ. of Groningen
\bibitem[Cannon et al.(2004)]{cmsc04} Cannon, J.~M., McClure-Griffiths, N.~M., Skillman, E.~D.,~\&
C\^ot\'e, S. 2004, AJ, in press
\bibitem[C\^{o}t\'{e} et al.(2000)]{cot00} C\^{o}t\'{e}, S., Carignan, C., \& Freeman, K.~C. 2000, \aj, 120, 3027
\bibitem[de Vaucouleurs et al.(1991)]{dev91} de Vaucouleurs, G., de Vaucouleurs, A., Corwin, H.~G., Buta, R.J., Paturel, G.,~\& Fouqu\'{e}, P. 1991 Third Reference Catalog of Bright Galaxies, Berlin:Springer-Verlag
\bibitem[Dolphin(2000)]{dol00} Dolphin, A.E., 2000, \apj, 531, 804
\bibitem[Elmegreen \& Elmegreen(1980)]{elm80} Elmegreen, D.~M.,~\& Elmegreen, B.~G. 1980 \aj, 85, 10 
\bibitem[Gallouet et al.(1975)]{gal75} Gallouet, L., Heidmann, N., \& Dampierre, F. 1975, \aaps, 19, 1 
\bibitem[Giovanelli \& Haynes(1988)]{gio88} Giovanelli R.,~\& Haynes M.~P. 1988, 
Galactic and Extragalactic Radio Astronomy, eds. Verschur G.~L.~\& Kellermann K.~I., 522
\bibitem[Hodge \& Miller(1995)]{hod95} Hodge, P.,~\& Miller, B.~W. 1995, \apj, 451, 176
\bibitem[Huchtmeier et al.(1981)]{huc81} Huchtmeier, W.~K., Seiradakis, J.~H.,~\& Materne, J., 1981, \aap, 102, 134
\bibitem[Huchtmeier \& Richter(1986)]{huc86} Huchtmeier, W.~K.,~\& Richter, O.~G. 1986, 
\aaps, 63, 323
\bibitem[Israel \& van Driel(1990)]{id90} Israel, F.~P.~\& van Driel, W.\ 1990, \aap, 236, 323 
\bibitem[Koribalski et al.(2004)]{kor04} Koribalski, B.~S. et al. 2004, \aj, in press
\bibitem[Massey et al.(2002)]{mas02} Massey, P., Hodge, P.~W., Holmes, S., Jacoby, G., King, N.~L., 
Olsen, K., Saha, A.,~\& Smith, C. AAS 199th meeting, Washington, DC, January 2002, 130.05
\bibitem[Mateo(1998)]{mat98} Mateo, M. 1998, \araa, 36, 435
\bibitem[Melisse \& Israel(1994)]{mel94} Melisse, J.~P.~M.,~\& Israel, F.~P. 1994, \aaps. 103, 391
\bibitem[Melotte(1926)]{mel26} Melotte, P.~J. 1926 \mnras, 86, 636
\bibitem[Miller(1996)]{mil96} Miller, B. 1996, \aj, 112, 991
\bibitem[Minniti \& Zijlstra(1996)]{min96} Minniti, D.,~\& Zijlstra, A.~A. 1996, \apj, 467, L13
\bibitem[Puche et al.(1992)]{pwbr92} Puche, D., Westpfahl, D., Brinks, E., \& Roy, J.\ 1992, \aj, 103, 1841 
\bibitem[Putman et al.(1998)]{put98} Putman, M.~E. et al. 1998, \nat, 394, 752
\bibitem[Roberts(1962)]{rob62} Roberts, M.~S. 1962, \aj, 67, 431
\bibitem[Roberts, Huntley \& van Albada(1979)]{rob79} Roberts, W.~W., Huntley, J.~M.,~\& van Albada, G.~D. 1979, \apj, 233, 67
\bibitem[Schlegel et al.(1998)]{sch98}Schlegel, D.~J., Finkbeiner,~\& D.~P., Davis, M. 1998, \apj, 500, 525
\bibitem[Shostak \& Skillman(1989)]{ss89} Shostak, G.~S.~\& Skillman, E.~D.\ 1989, \aap, 214, 33 
\bibitem[Skillman(1996)]{ski96} Skillman, E.~D. 1996, in ASP Conf. Ser. 106, The Minnesota Lectures on 
Extragalactic Neutral Hydrogen, ed. E.~D. Skillman, (San Francisco: ASP), 208
\bibitem[Stil \& Israel(2002)]{si02} Stil, J.~M.~\& Israel, F.~P.\ 2002, \aap, 392, 473 
\bibitem[Taylor et al.(1995)]{tay95} Taylor, C.~L., Brinks, E., Grashuis, R.~M.,~\& Skillman, E.~D. 1995, \apjs, 99, 427 
\bibitem[Taylor \& Klein(2001)]{tay01} Taylor, C.~L.,~\& Klein, U. 2001, \aap, 366, 811
\bibitem[Taylor, Kobulnicky, \& Skillman(1998)]{tks98} Taylor, C.~L., Kobulnicky, H.~A., \& Skillman, E.~D.\ 1998, \aj, 116, 2746 
\bibitem[van den Bergh(1994)]{van94} van den Bergh, S. 1994, \aj, 107, 1328
\bibitem[van den Bergh(2000)]{van00} van den Bergh, S. 2000, The Galaxies of the Local Group (Cambridge, U.K.;Cambridge University Press)
\bibitem[Venn et al.(2003)]{ven03} Venn, K.~A., Tolstoy, E., Kaufer, A., Skillman, E.~D., Clarkson, S.M., 
Smartt, S.~J., Lennon, D.~J.,~\& Kudritzki, R.~P. 2003, \aj, 126, 1326 
\bibitem[Wilcots \& Miller(1998)]{wm98} Wilcots, E.~M.~\& Miller, B.~W.\ 1998, \aj, 116, 2363 
\bibitem[Wolf(1910)]{wol10} Wolf, M. 1910, Astron. Nachr., 183, 187
\end{thebibliography}
\end{document}